\documentclass[aps,preprint,showpacs,preprintnumbers,amsmath,amssymb]{revtex4}

\begin{document}

\title{An integral containing the product of four Bessel functions}

\author{Cosmin Crucean \\
{\normalsize  \it West University of Timi\c soara,}\\
{\normalsize  \it V. Parvan Ave. 4 RO-300223 Timi\c soara,  Romania}}

\begin{abstract}
Mellin transform is used to evaluate an integral involving the product of four Bessel functions and a power. Using this method the result is obtained in terms of generalized hypergeometric functions $_{6}F_{5}$.
\end{abstract}


\maketitle

\section{Introduction}
The integral that we want to solve in this paper is of the type:
\begin{equation}
\int_{0}^{\infty} dx \,x^{\mu}J_{\alpha}(ax)J_{\beta}(ax)J_{\gamma}(bx)J_{\delta}(bx).\label{b}
\end{equation}

In our analysis the two cases: $a>b$ and $b>a$ will be discussed. Our intention is to evaluate the above integral using the Mellin transform for representing relation (\ref{b}) as a contour integral in the complex plane, whose integrand will contain ratios of gamma functions. The evaluation that will be presented here follow the method used in \cite{GT}, based on Mellin transforms, that can be used for evaluating the integrals of this type.

The paper is organized as follows: in Section 2 we introduce the Mellin transform for a Bessel function as in \cite{GT} and we write the integral as a contour integral in the complex plane. In Section 3 we present the main steps that help us to arrive at the final result. In Section 4 we point out the possible applications in the scattering theory on an expanding universe.

\section{Mellin transform}
The integral (\ref{b}) can be brought in the form \cite{GT}:
\begin{equation}
a^{-\mu-1}\int_{0}^{\infty} d\eta \,\eta^{\mu}J_{\alpha}(\eta)J_{\beta}(\eta)J_{\gamma}(\tau\eta)J_{\delta}(\tau\eta)=a^{-\mu-1}
I_{\alpha\beta\gamma\delta}(\mu,\tau),\label{c}
\end{equation}
where we use the notations $\eta=ax$ and $\tau=b/a$.

The solution that we obtain is valid for a range of parameters consistent with the convergence of the integral from relation (\ref{c}). A study of this expression reveals that the integral converges when:
\begin{eqnarray}
&&Re(\mu)<2, \nonumber\\
&&Re(\mu+\alpha+\beta+\gamma+\delta+1)>0 \label{asim}
\end{eqnarray}
where the first expression is obtained by considering the behavior of the integrand of (\ref{c}) as $\eta$ approaches infinity and the second expression is obtained by considering the behavior of the integrand as $\eta$ approaches to zero \cite{AS},\cite{GR}.

The integral that interest us can be brought in the following form:
\begin{equation}
I_{\alpha\beta\gamma\delta}(\mu,\tau)=\int_{0}^{\infty}d\eta \,\eta^{\mu}M_{1}(\eta)M_{2}(\tau\eta)\label{int}
\end{equation}
where $M_{1}(\eta)$ and $M_{2}(\tau\eta)$ are functions whose Mellin transform can be obtained as ratios of products of gamma functions.
Let us start by considering the Mellin transform of (\ref{c}). The Mellin transform of a function $M(\eta)$ will be in our notations $\mathcal{M}(s)$, and will have the following expression:
\begin{equation}
\mathcal{M}(s)=\int_{0}^{\infty}d\eta \,\eta^{s-1}M(\eta).\label{mel}
\end{equation}
With the observation that the original function is recovered from its Mellin transform evaluating the contour integral:
\begin{equation}
M(\eta)=\frac{1}{2\pi i}\int_{\lambda-i\infty}^{\lambda+i\infty}ds \,\eta^{-s}\mathcal{M}(s),\label{inv}
\end{equation}
where $\lambda$ is a small positive constant. The Mellin transform of $I_{\alpha\beta\gamma\delta}(\mu,\tau)$ will be \cite{GT}:
\begin{equation}
\mathcal{I_{\alpha\beta\gamma\delta}}(\mu,s)=\int_{0}^{\infty}d\tau \, \tau^{s-1}I_{\alpha\beta\gamma\delta}(\mu,\tau).\label{mi}
\end{equation}
Now substitute (\ref{int}) in equation (\ref{mi}) and using equation (\ref{mel}) we obtain:
\begin{equation}
\mathcal{I_{\alpha\beta\gamma\delta}}(\mu,s)=\mathcal{M}_{1}(\mu-s+1)\mathcal{M}_{2}(s).\label{mm}
\end{equation}
It not difficult now to use the Mellin inversion formula given by (\ref{inv}), for obtain an alternative expression for (\ref{c}):
\begin{equation}
I_{\alpha\beta\gamma\delta}(\mu,\tau)=\frac{1}{2\pi i}\int_{\lambda-i\infty}^{\lambda+i\infty}ds \,\tau^{-s}\mathcal{M}_{1}(\mu-s+1)\mathcal{M}_{2}(s).\label{mf}
\end{equation}

In our case the functions whose Mellin transform can be expressed as ratios of gamma functions are:
\begin{eqnarray}
&&M_{1}=J_{\alpha}(\eta)J_{\beta}(\eta),\nonumber\\
&&M_{2}=J_{\gamma}(\tau\eta)J_{\delta}(\tau\eta).
\end{eqnarray}
The required Mellin transforms are:
\begin{eqnarray}
&&\mathcal{M}_{1}(\mu-s+1)=
\frac{\Gamma(s-\mu)\Gamma(\frac{\alpha+\beta+\mu-s+1}{2})}{2^{s-\mu}\Gamma(\frac{-\alpha+\beta+s-\mu+1}{2})
\Gamma(\frac{\alpha+\beta+s-\mu+1}{2})\Gamma(\frac{\alpha-\beta+s-\mu+1}{2})}\,\,\,,\nonumber\\
&&\mathcal{M}_{2}(s)=
\frac{\Gamma(1-s)\Gamma(\frac{\gamma+\delta+s}{2})}{2^{1-s}\Gamma(\frac{-\gamma+\delta-s+2}{2})
\Gamma(\frac{\gamma+\delta-s+2}{2})\Gamma(\frac{\gamma-\delta-s+2}{2})}.\label{m12}
\end{eqnarray}
Replacing (\ref{m12}) in (\ref{mf}) we obtain:
\begin{eqnarray}
I_{\alpha\beta\gamma\delta}(\mu,\tau)&=&\frac{1}{2\pi i}\int_{\lambda-i\infty}^{\lambda+i\infty}ds \,\tau^{-s}
\frac{\Gamma(s-\mu)\Gamma(\frac{\alpha+\beta+\mu-s+1}{2})}{2^{s-\mu}\Gamma(\frac{-\alpha+\beta+s-\mu+1}{2})
\Gamma(\frac{\alpha+\beta+s-\mu+1}{2})\Gamma(\frac{\alpha-\beta+s-\mu+1}{2})}\,\,\,\nonumber\\
&&\times\frac{\Gamma(1-s)\Gamma(\frac{\gamma+\delta+s}{2})}{2^{1-s}\Gamma(\frac{-\gamma+\delta-s+2}{2})
\Gamma(\frac{\gamma+\delta-s+2}{2})\Gamma(\frac{\gamma-\delta-s+2}{2})}.
\end{eqnarray}

Making the coordinate transformations $s=2u$ and $c=\lambda/2$ as in \cite{GT} and using the duplication formula for gamma functions \cite{AS}
\begin{equation}
\Gamma(2z)=(2\pi)^{-1/2}2^{2z-1/2}\Gamma(z)\Gamma(z+1/2)
\end{equation}
we obtain the following result:
\begin{eqnarray}
I_{\alpha\beta\gamma\delta}(\mu,\tau)&=&\frac{-i}{4\pi^{2}}\int_{c-i\infty}^{c+i\infty}du\, \tau^{-2u}
\frac{\Gamma(u-\frac{\mu}{2})\Gamma(u+\frac{-\mu+1}{2})\Gamma(\frac{\alpha+\beta+\mu+1}{2}-u)}{\Gamma(\frac{-\alpha+\beta-\mu+1}{2}+u)
\Gamma(\frac{\alpha+\beta-\mu+1}{2}+u)\Gamma(\frac{\alpha-\beta-\mu+1}{2}+u)}\,\,\,\nonumber\\
&&\times\frac{\Gamma(1-u)\Gamma(\frac{1}{2}-u)\Gamma(\frac{\gamma+\delta}{2}+u)}{\Gamma(\frac{-\gamma+\delta+2}{2}-u)
\Gamma(\frac{\gamma+\delta+2}{2}-u)\Gamma(\frac{\gamma-\delta+2}{2}-u)}.\label{f}
\end{eqnarray}
The above integral is a contour integral in complex plane. As one know the poles of the gamma function occur when their arguments are zero or negative integers. Then the integral can be evaluated using the fact that the poles of the numerator appears for the above mentioned values. Since there are an infinite number of such poles inside the contour an infinite series will be obtained when the residue of these poles are summed.

\section{Evaluation of the integral}
The integral from relation (\ref{f}) will be evaluated using the complex-variable technique. Each gamma function from numerator has a pole when the argument equals zero or is an negative integer. The poles of the integrand are:
\begin{eqnarray}
u&=&\frac{\mu}{2}-k\,;\quad
u=\frac{\mu-1}{2}-k\,; \nonumber\\
u&=&-k-\frac{(\gamma+\delta)}{2}.\label{pol1}
\end{eqnarray}
and
\begin{eqnarray}
u=1+k\,; \quad u=\frac{1}{2}+k\,;\nonumber\\
u=\frac{\mu+\alpha+\beta+1}{2}+k\,,\label{pol2}
\end{eqnarray}
with $k=0,1,2,3...$.
For the following calculations we assume that the values of parameters $\mu,\alpha,\beta,\gamma,\delta$ are such that a value for $c$ can be found for which the poles from expression (\ref{pol1}) lie to the left of line $u=c$ while the poles given by (\ref{pol2}) lie to the right of line $u=c$.

Let us begin with the case $a>b\,\,\,(\tau<1)$, when the poles represented by (\ref{pol1}) give contributions. For evaluating the integral (\ref{f}) we use the following relation:
\begin{equation}
\oint_{C_{1}}dz H(z)\Gamma(z)=2\pi i \sum_{k=0}^{\infty}\frac{(-1)^{k}}{k!}H(-k),\label{hk}
\end{equation}
where $H(z)$ is a analytic function without singularities in the left half-plane. Using (\ref{hk}) we find that the final result is:
\begin{eqnarray}
&&I_{\alpha\beta\gamma\delta}(\mu,\tau)=\frac{1}{2\pi}\biggl[\sum_{k=0}^{\infty}\frac{(-1)^{k}}{k!}\,\frac{\tau^{-\mu+2k}\,
\Gamma(\frac{1}{2}-k)\Gamma(k+\frac{-\mu+1}{2})\Gamma(\frac{\alpha+\beta+1}{2}+k)}{\Gamma(\frac{-\alpha+\beta+1}{2}-k)
\Gamma(\frac{\alpha+\beta+1}{2}-k)\Gamma(\frac{\alpha-\beta+1}{2}-k)}\,\,\,\nonumber\\
&&\times\frac{\Gamma(1-\frac{\mu}{2}+k)\Gamma(\frac{\gamma+\delta+\mu}{2}-k)}{\Gamma(\frac{-\gamma+\delta-\mu+2}{2}+k)
\Gamma(\frac{\gamma+\delta-\mu+2}{2}+k)\Gamma(\frac{\gamma-\delta-\mu+2}{2}+k)}+\sum_{k=0}^{\infty}\frac{(-1)^{k}}{k!}
\,\frac{\tau^{-\mu+2k+1}}{\Gamma(\frac{-\alpha+\beta}{2}-k)}\nonumber\\
&&\times\frac{\Gamma(-\frac{1}{2}-k)\Gamma(k+\frac{-\mu+2}{2})\Gamma(\frac{\alpha+\beta+2}{2}+k)\Gamma(\frac{3-\mu}{2}+k)
\Gamma(\frac{\gamma+\delta+\mu-1}{2}-k)}{\Gamma(\frac{\alpha+\beta}{2}-k)\Gamma(\frac{\alpha-\beta}{2}-k)
\Gamma(\frac{-\gamma+\delta-\mu+3}{2}+k)
\Gamma(\frac{\gamma+\delta-\mu+3}{2}+k)\Gamma(\frac{\gamma-\delta-\mu+3}{2}+k)}\,\nonumber\\
&&+\sum_{k=0}^{\infty}\frac{(-1)^{k}}{k!}\frac
{\tau^{\gamma+\delta+2k}\,\Gamma(\frac{\gamma+\delta+1}{2}+k)\Gamma(\frac{2+\gamma+\delta}{2}+k)
\Gamma(\frac{-\gamma-\delta-\mu}{2}-k)}{\Gamma(\frac{\alpha+\beta-\mu-\gamma-\delta+1}{2}-k)
\Gamma(\frac{-\alpha+\beta-\mu-\gamma-\delta+1}{2}-k)\Gamma(\frac{\alpha-\beta-\mu-\gamma-\delta+1}{2}-k)}\nonumber\\
&&\times\frac{\Gamma(\frac{-\gamma+\delta-\mu+1}{2}-k)
\Gamma(\frac{\gamma+\delta+\alpha+\beta+\mu+1}{2}+k)}{\Gamma(\gamma+k+1)\Gamma(\delta+k+1)
\Gamma(\gamma+\delta+k+1)}\biggl].\label{c1}
\end{eqnarray}
This result can be written now in terms of generalized hypergeometric functions, using the following relation:
\begin{eqnarray}
\sum_{k=0}^{\infty}\frac{z^{k}}{k!}\frac{\prod^{N_{g}}_{i=1}\,\Gamma(k+g_{i})\prod^{N_{m}}_{i=1}\Gamma(m_{i}-k)}
{\prod^{N_{q}}_{i=1}\Gamma(k+q_{i})\prod^{N_{p}}_{i=1}\Gamma(p_{i}-k)}=\frac{\prod^{N_{g}}_{i=1}\Gamma(g_{i})
\prod^{N_{m}}_{i=1}\Gamma(m_{i})}
{\prod^{N_{q}}_{i=1}\Gamma(q_{i})\prod^{N_{p}}_{i=1}\Gamma(p_{i})}\nonumber\\
_{N_{g}+N{p}}\,F_{N_{m}+N_{q}}(g_{1},g_{2},...,g_{N_{g}},-p_{1}+1,-p_{2}+1,...-p_{N_{p}}+1;q_{1},q_{2},\nonumber\\
...,q_{N_{q}}, -m_{1}+1,-m_{2}+2,...,-m_{N_{m}}+1;(-1)^{N_{m}-N_{p}}z).\label{h}
\end{eqnarray}
Using equation (\ref{h}) and the following relations \cite{AS},\cite{GR},
\begin{eqnarray}
&&\Gamma(1/2)=\pi^{1/2},\quad \Gamma(-1/2)=-2\pi^{1/2}, \nonumber\\
&&\Gamma(z)\Gamma(z+1/2)=2^{1-2z}\pi^{1/2}\Gamma(2z)\label{g12}
\end{eqnarray}
the final form of equation (\ref{c1}) is:
\begin{eqnarray}
&&I_{\alpha\beta\gamma\delta}(\mu,\tau)=\frac{2^{-1+\mu}\,\tau^{-\mu}\,\Gamma(\frac{-\mu+1}{2})\Gamma(\frac{\alpha+\beta+1}{2})
\Gamma(\frac{\gamma+\delta+\mu}{2})}{\Gamma(\frac{\alpha+\beta+1}{2})\Gamma(\frac{\alpha-\beta+1}{2})
\Gamma(\frac{-\alpha+\beta+1}{2})
\Gamma(\frac{-\gamma+\delta-\mu+2}{2})
\Gamma(\frac{\gamma+\delta-\mu+2}{2})}\nonumber\\
&&\times\frac{1}{\Gamma(\frac{\gamma-\delta-\mu+2}{2})}\,_{6}F_{5}\biggl(\frac{\alpha+\beta+1}{2},\frac{1-\mu}{2},\frac{2-\mu}{2},
\frac{\alpha-\beta+1}{2},\frac{-\alpha-\beta+1}{2},\nonumber\\
&&\frac{-\alpha+\beta+1}{2};\frac{-\gamma+\delta-\mu+2}{2},\frac{\gamma+\delta-\mu+2}{2},\frac{\gamma-\delta-\mu+2}{2},
\frac{1}{2},\nonumber\\
&&\frac{-\gamma-\delta-\mu+2}{2};\tau^{2}\biggl)-\frac{2^{-2+\mu}\,\tau^{-\mu+1}\,\Gamma(2-\mu)\Gamma(\frac{\alpha+\beta+2}{2})
\Gamma(\frac{\gamma+\delta+\mu-1}{2})}{\Gamma(\frac{\alpha+\beta}{2})\Gamma(\frac{\alpha-\beta}{2})\Gamma(\frac{-\alpha+\beta}{2})
\Gamma(\frac{-\gamma+\delta-\mu+3}{2})\Gamma(\frac{\gamma+\delta-\mu+3}{2})}\nonumber\\
&&\times\frac{1}{\Gamma(\frac{\gamma-\delta-\mu+3}{2})}\,_{6}F_{5}\biggl(\frac{\alpha+\beta+2}{2},\frac{2-\mu}{2},\frac{3-\mu}{2},
\frac{\alpha-\beta+2}{2},\frac{-\alpha-\beta+2}{2},\nonumber\\
&&\frac{-\alpha+\beta+2}{2};\frac{-\gamma+\delta-\mu+3}{2},\frac{\gamma+\delta-\mu+3}{2},\frac{\gamma-\delta-\mu+3}{2},
\frac{3}{2},\nonumber\\
&&\frac{-\gamma-\delta-\mu+3}{2};\tau^{2}\biggl)+\frac{2^{\mu}\,\tau^{\gamma+\delta}\,
\Gamma(\frac{-\gamma-\delta-\mu}{2})\Gamma(\frac{\alpha+\beta+\mu+\gamma+\delta+1}{2})
\Gamma(\frac{\gamma+\delta+1}{2})}{\Gamma(\frac{\alpha+\beta-\gamma-\delta-\mu+1}{2})
\Gamma(\frac{\alpha-\beta-\gamma-\delta-\mu+1}{2})\Gamma(\frac{-\alpha+\beta-\gamma-\delta-\mu+1}{2})}\nonumber\\
&&\times\frac{1}{\Gamma(\gamma+1)\Gamma(\delta+1)\Gamma(\gamma+\delta+1)}\,
_{6}F_{5}\biggl(\frac{\alpha+\beta+\mu+\gamma+\delta+1}{2},\frac{\gamma+\delta+1}{2},\nonumber\\
&&\frac{\gamma+\delta+2}{2},\frac{\alpha-\beta+\mu+\gamma+\delta+1}{2},\frac{-\alpha-\beta+\mu+\gamma+\delta+1}{2},\nonumber\\
&&\frac{-\alpha+\beta+\mu+\gamma+\delta+1}{2};\delta+1,\gamma+\delta+1,\gamma+1,
\frac{\mu+\gamma+\delta+2}{2},\nonumber\\
&&\frac{\mu+\gamma+\delta+1}{2};\tau^{2}\biggl).\label{fc1}
\end{eqnarray}

When $\tau>1$ the contour is completed in the right half-pane. In this case the poles that give contributions are given in equation (\ref{pol2}). Now the sign of equation (\ref{f}) must be changed because the contour is traversed in the opposite direction. The contribution due to the poles will be evaluated this time using:
\begin{equation}
\oint_{C_{2}}dz H(z)\Gamma(-z)=-2\pi i\sum_{k=0}^{\infty}\frac{(-1)^{k}}{k!}H(k),\label{hz}
\end{equation}
where $H(z)$ is a analytic function free of singularities in the right half-plane. Using (\ref{hz}) to evaluate (\ref{f}) we finally obtain:
\begin{eqnarray}
&&I_{\alpha\beta\gamma\delta}(\mu,\tau)=\frac{1}{2\pi}\biggl[\sum_{k=0}^{\infty}\frac{(-1)^{k}}{k!}\,
\frac{\tau^{-(\alpha+\beta+\mu+2k+1)}\,
\Gamma(k+\frac{\alpha+\beta+1}{2})\Gamma(\frac{\alpha+\beta+2}{2}+k)}{\Gamma(\beta+1+k)\Gamma(\alpha+\beta+1+k)
\Gamma(\alpha+1+k)}\,\,\,\nonumber\\
&&\times\frac{\Gamma(\frac{-\alpha-\beta-\mu}{2}-k)\Gamma(\frac{1-\alpha-\beta-\mu}{2}-k)
\Gamma(\frac{\gamma+\delta+\alpha+\beta+\mu+1}{2}+k)}{\Gamma(\frac{-\gamma+\delta-\alpha-\beta-\mu+1}{2}-k)
\Gamma(\frac{\gamma+\delta-\alpha-\beta-\mu+1}{2}-k)\Gamma(\frac{\gamma-\delta-\alpha-\beta-\mu+1}{2}-k)}+
\sum_{k=0}^{\infty}\frac{(-1)^{k}}{k!}\nonumber\\
&&\times\frac{\tau^{-2k-1}\,\Gamma(\frac{1-\mu}{2}+k)\Gamma(\frac{1}{2}-k)\Gamma(\frac{\alpha+\beta+\mu}{2}-k)
\Gamma(\frac{2-\mu}{2}+k)}{\Gamma(\frac{-\mu-\alpha+\beta+2}{2}+k)\Gamma(\frac{\alpha+\beta-\mu+2}{2}+k)
\Gamma(\frac{\alpha-\beta-\mu+2}{2}+)\Gamma(\frac{-\gamma+\delta+1}{2}-k)
\Gamma(\frac{\gamma+\delta+1}{2}-k)}\,\nonumber\\
&&\times\frac{\Gamma(\frac{\gamma+\delta+1}{2}+k)}{\Gamma(\frac{\gamma-\delta+1}{2}-k)}+\sum_{k=0}^{\infty}\frac{(-1)^{k}}{k!}\frac
{\tau^{-2-2k}\,\Gamma(\frac{\alpha+\beta+\mu-1}{2}-k)\Gamma(\frac{2+\gamma+\delta}{2}+k)
\Gamma(\frac{2-\mu}{2}+k)}{\Gamma(\frac{\alpha+\beta-\mu+3}{2}+k)\Gamma(\frac{-\alpha+\beta-\mu+3}{2}+k)}\nonumber\\
&&\times\frac{\Gamma(\frac{3-\mu}{2}+k)\Gamma(-k-\frac{1}{2})}{\Gamma(\frac{\alpha-\beta-\mu+3}{2}+k)
\Gamma(\frac{-\gamma+\delta}{2}-k)\Gamma(\frac{\gamma+\delta}{2}-k)\Gamma(\frac{\gamma-\delta}{2}-k)}\biggl].
\end{eqnarray}
For obtaining the final result we must use now (\ref{g12}) and (\ref{h}):
\begin{eqnarray}
&&I_{\alpha\beta\gamma\delta}(\mu,\tau)=\frac{2^{\mu}\,\tau^{-(\alpha+\beta+\mu+1)}\,
\Gamma(\frac{-\alpha-\beta-\mu}{2})\Gamma(\frac{\alpha+\beta+1}{2})
\Gamma(\frac{\alpha+\beta+\gamma+\delta+\mu+1}{2})}{\Gamma(\alpha+1)\Gamma(\beta+1)
\Gamma(\alpha+\beta+1)\Gamma(\frac{-\gamma+\delta-\alpha-\beta-\mu+1}{2})
\Gamma(\frac{\gamma+\delta-\alpha-\beta-\mu+1}{2})}\nonumber\\
&&\times\frac{1}{\Gamma(\frac{\gamma-\delta-\alpha-\beta-\mu+1}{2})}\,_{6}F_{5}\biggl(\frac{\alpha+\beta+1}{2},
\frac{\alpha+\beta+2}{2},\frac{\gamma+\delta+\alpha+\beta+\mu+1}{2},\nonumber\\
&&\frac{\gamma-\delta+\alpha+\beta+\mu+1}{2},\frac{-\gamma-\delta+\alpha+\beta+\mu+1}{2},\nonumber\\
&&\frac{-\gamma+\delta+\alpha+\beta+\mu+1}{2};\beta+1,\delta+\beta+1,\alpha+1,
\frac{\alpha+\beta+\mu+2}{2},\nonumber\\
&&\frac{\alpha+\beta+\mu+1}{2};\tau^{-2}\biggl)+\frac{2^{-1+\mu}\,\tau^{-1}\,\Gamma(\frac{1-\mu}{2})\Gamma(\frac{\alpha+\beta+\mu}{2})
\Gamma(\frac{\gamma+\delta+1}{2})}{\Gamma(\frac{\alpha+\beta-\mu+2}{2})\Gamma(\frac{\alpha-\beta-\mu+2}{2})\Gamma(\frac{\delta-\gamma+1}{2})
\Gamma(\frac{\gamma+\delta+1}{2})\Gamma(\frac{\gamma-\delta+1}{2})}\nonumber\\
&&\times\frac{1}{\Gamma(\frac{\beta-\alpha-\mu+2}{2})}\,_{6}F_{5}\biggl(\frac{1-\mu}{2},\frac{2-\mu}{2},
\frac{\gamma+\delta+1}{2},\frac{\gamma-\delta+1}{2},\frac{-\gamma-\delta+1}{2},\nonumber\\
&&\frac{-\gamma+\delta+1}{2};\frac{-\alpha+\beta-\mu+2}{2},\frac{\alpha+\beta-\mu+2}{2},\frac{\alpha-\beta-\mu+2}{2},
\nonumber\\
&&\frac{-\alpha-\beta-\mu+2}{2},\frac{1}{2};\tau^{-2}\biggl)-\frac{2^{-2+\mu}\,\tau^{-2}\,
\Gamma(\frac{2-\mu}{2})\Gamma(\frac{\alpha+\beta+\mu-1}{2})\Gamma(\frac{\gamma+\delta+2}{2})}
{\Gamma(\frac{\alpha+\beta-\mu+3}{2})
\Gamma(\frac{\alpha-\beta-\mu+3}{2})\Gamma(\frac{-\alpha+\beta-\mu+3}{2})}\nonumber\\
&&\times\frac{1}{\Gamma(\frac{\gamma-\delta}{2})\Gamma(\frac{\delta+\gamma}{2})\Gamma(\frac{-\gamma+\delta}{2})}\,
_{6}F_{5}\biggl(\frac{2-\mu}{2},\frac{3-\mu}{2},\frac{\gamma+\delta+2}{2},\frac{\gamma-\delta+2}{2}
,\frac{2-\gamma-\delta}{2},\nonumber\\
&&\frac{-\gamma+\delta+2}{2};\frac{\beta-\alpha-\mu+3}{2},\frac{\alpha+\beta-\mu+3}{2},\frac{\alpha-\beta-\mu+3}{2},
\frac{-\alpha-\beta-\mu+3}{2},\nonumber\\
&&\frac{3}{2};\,\tau^{-2}\biggl).\label{fc2}
\end{eqnarray}
This is the result of the integral in the case $\tau>1$.

\section{Conclusion}
We presented here a method based on Mellin transform to evolve one integral that contain the product of four Bessel functions with two different arguments.
The integrals presented here could be of interest in the perturbative  QED on de Sitter space and in the study of Feynman propagators. It is known that in de Sitter case the solutions of the free field equations for Dirac and scalar fields are given in terms of Hankel functions which can be expressed with the help of Bessel function. Also in the present literature the scattering amplitudes on curved space seems to receive a little attention. An explicit calculation show that the scattering amplitudes in the first order of perturbation theory, imply integrals with products of two Bessel functions. Hoverer in second  and fourth order of the perturbation theory the situation seems to be more complicated with integrals that contain products of multiple Bessel functions.

\end{document}